\documentclass[12pt]{article}
\usepackage{amsmath,amssymb}
\usepackage{graphicx,color}
\numberwithin{equation}{section}
\usepackage{cite}
\usepackage{bm}
\usepackage{dcolumn}
\newcommand{\be}{\begin{equation}}
\newcommand{\bea}{\begin{eqnarray}}
\newcommand{\eea}{\end{eqnarray}}
\newcommand{\ba}{\begin{array}}
\newcommand{\ea}{\end{array}}
\newcommand{\ee}{\end{equation}}

\expandafter\ifx\csname mathbbm\endcsname\relax

\else
simple linear group

\fi

\numberwithin{equation}{section}
\newcommand\email[4]{#1@#2.#3.#4}
\newcommand\Email[5]{#1@#2.#3.#4.#5}

\newcommand\atmp[3]{Adv. Theor. Math. Phys. {\bf #1}~(#2)~#3}

\newcommand\jmp[3]{J. Math. Phys.{\bf {#1}}~({#2})~{#3}}

\newcommand\jhep[3]{JHEP {\bf {#1}}~({#2})~{#3}}

\newcommand\npb[3]{Nucl.~Phys. {\bf B{#1}}~({#2})~{#3}}

\newcommand\plb[3]{Phys. Lett. {\bf B{#1}}~({#2})~{#3}}

\begin{document}
\title{
\begin{flushright}
\small IPM/P-2008/014
\end{flushright}
\addvspace{1cm}
{\bf  On Six Dimensional Fundamental Superstrings as Holograms}}

\author{
Mohsen Alishahiha$^a$\thanks{\Email{alishah}{theory}{ipm}{ac}{ir}}  
~~and~~ 
Subir Mukhopadhyay$^b$\thanks{\email{subir}{iopb}{res}{in}}\\ \\
\small ${}^a$ {\em Institute for Studies in Theoretical Physics and Mathematics,}\\ 
\small {\em P.O. Box 19395-5531, Tehran, Iran.} 
\\ \\
\small ${}^b$ {\em Institute of Physics, Bhubaneswar 751~005, India.} 
 }
\date{}
\maketitle
\begin{abstract}
\small
\noindent In this note we discuss a possible holographic dual of the two dimensional conformal field theory associated with the world-sheet of a macroscopic superstring in a compactification on four-torus. We assume the near horizon geometry of the black string has symmetries of $AdS_3\times S^3\times T^4$  and construct a sigma model in the bulk. Analyzing the symmetries of the bulk theory and comparing them with those of the CFT in a special light-cone gauge we find agreement between global symmetries.  Due to non-standard gauge realization it is not clear how affine symmetries can be realized. 
\end{abstract}

\section{Introduction}
The world-sheet theory living on a macroscopic fundamental superstring is described by a two dimensional superconformal field theory
with central charge $(c_L,c_R)=(12p,12p)$ for type II and $(24p,12p)$ for heterotic string theories, where $p$ is the
number of stretched, coincident fundamental strings. Recently it has been proposed that there may exist a holographic dual to these world-sheet theories\cite{Dabh,johnson,strominger,kraus}
(See \cite{{Hung:2007su},{Alishahiha:2007ap}} for related topic and see \cite{Castro:2008ne}
for recent review). 
There are a couple of reasons for such expectations\cite{strominger}. Since the macroscopic string appears as a defect in the supergravity, the Maldacena duality suggests that the world-sheet theory living on the defect may be dual to the string theory living on its near horizon limit. Since in low energy the world-sheet theory is described by a CFT the dual geometry should contain an AdS$_3$. On the other hand, since at the core the curvature blows up, contributions of different orders of $\alpha^\prime$ become equally important one should look for an exact CFT description. Moreover, since the string coupling vanishes at the core this situation is suitable for a CFT analysis. The second reason comes from the exact matching of counting of microscopic states of a string wrapped on a circle and the macroscopic entropy of the resulting two-charge small black hole calculated a la Wald\cite{{Dabholkar:2004yr},{Dabholkar:2005dt},{Dabholkar:2005by}}. This agreement turns out to hold at all orders of $\alpha^\prime$ and it has been suggested in \cite{strominger} that since the leading term in the entropy does not contribute it is hard to understand how this exact matching occurs unless there is a dual CFT description.

In this spirit the string theory on the near horizon limit of a small black string is considered in\cite{Dabh,johnson,strominger,kraus} for a five-torus compactification. Compactifying on one more circle down to four dimension one obtains small black hole which, after taking $\alpha^\prime$ corrections into account, leads to an AdS$_2\times$S$^2$ near horizon geometry. From five dimensional point of view, AdS$_2$ with a circle fibered on it leads to an AdS$_3$ with some global identifications. Going to the covering space this becomes an AdS$_3\times$S$^2$ geometry. In five dimensions, using supersymmtrized $R^2$ corrections, 
an explicit solution which interpolates between $AdS_3\times S^2$ 
near horizon geometry and asymptotic flat space has been exhibited 
in \cite{cdkl} (See \cite{{Alishahiha:2007nn},{Cvitan:2007hu}} for more discussions on this model). 

A few studies about the holographic dual have appeared recently for five-torus compactification. In \cite{strominger} the analysis of five dimensional off-shell supergravity shows the near horizon geometry $AdS_3\times S^2$ has appropriate killing spinors and bosonic isometries so as to give 
$Osp(4^*|4)$ super-isometry group associated with sixteen supercharges; they also discussed possible affine extensions. 
\cite{Dabh} analyzed a sigma model on $AdS_3\times S^2$
and compare the symmetries with those of the free world-sheet theory describing the dynamics of the macroscopic string. They showed that the global symmetries coming from the sigma model agrees with those of the hologram only in a specific light-cone gauge.

In all these studies the possibility of extending this analysis to higher dimension has been discussed and the possible superisometry group has been enlisted.
Though an explicit solution with higher derivative corrections in other dimensions is missing,
on general ground one can argue that adding bosonic Gauss-Bonnet can stretch the horizon leading to non-singular near horizon geometry of the small black hole \cite{sen2}. We note, however, that having got 
a stretched horizon with $AdS_2/AdS_3$ near horizon geometry does not necessarily mean that there is a solution in 
entire space time
which interpolates between this near horizon geometry and an asymptotic flat solution. In fact, a priory it is not obvious that such 
a solution exist. But on the basis of the arguments in \cite{strominger,Dabh,kraus}, one can expect that similar situation will prevail in other dimensions as well.

Since the theory depends crucially on dimensions it is not obvious how the five dimensional model studied so far can be generalized to higher dimensions. In the present paper we consider the study of the holographic dual for macroscopic string compactfied on a four-torus. A qualitative discussion of symmetry groups for
other dimensional compactification has already been appeared in
\cite{Dabh,strominger, kraus}.
Following suggestions of \cite{strominger, Dabh} 
we assume that the near horizon
geometry has the symmetries of $AdS_3\times S^3$, in analogy
with five dimensional case. An intuitive argument supporting this assumption comes from the fact that Gauss-Bonnet action
can indeed stretch the horizon leading $AdS_3/AdS_2$ near horizon geometry.

We begin with the two dimensional world-sheet CFT 
describing the world-sheet dynamics of the macroscopic string as the hologram. On the bulk side 
we consider a sigma model that describes the corresponding 
string theory on $AdS_3\times S^3$ along with a four-torus 
and compare the corresponding symmetries, both bosonic and 
fermionic with those of the hologram. Let us note that our work involves eight supercharges in each chiral supersymmetric sector and  so is different from the extensive studies on the string theory on $AdS_3\times S^3$ that corresponds to $D1/D5$ system appeared during the 
last decade, involving four supercharges in each chiral supersymmetric sector. 
We find as in \cite{Dabh} that one has to choose non trivial
light-cone gauge on the CFT side, in order to match some of 
the global symmetries, which makes the realization of corresponding affine extensions difficult. 

The plan of the paper is as follows. In section 2 we discuss the
hologram, namely the two dimensional CFT. In section 3 we 
construct the sigma model and analyze its symmetries and compare
them with that obtained in the hologram picture. Finally
we conclude in section 4 with a discussion.

\section{Hologram}

In this section we discuss the hologram, namely the CFT describing the world-sheet of a macroscopic superstring toroidally compactified on a four-torus in Green-Schwarz formulation. We begin with a review of the generators of symmetry transformations and the corresponding algebra in a light-cone gauge along a compact direction, which will also help setting up the notations. Then, following 
\cite{Dabh} we introduce a slightly different light-cone gauge and discuss the consequent modification of the algebra of the symmetry generators. 

Let $X^0 \dots X^9$ denote the target space coordinates where we denote the radius of $X^9$ circle by $R$. The mode expansions of non-compact bosons ($X^i, i=0,...,5$) and compact bosons ($X^m, m=6,..,9$) are given by
\begin{gather}
X^i = x^i + \alpha^\prime p^i \tau + i\sqrt{\frac{\alpha^\prime}{2}} \sum\limits_{n\neq 0} \frac{1}{n} [ \alpha^i_n e^{-2in(\tau -\sigma)} + {\tilde\alpha}^i_n e^{-2in(\tau +\sigma)} ], \label{modesnoncompact}\\
X^m = x^m + \frac{\alpha^\prime q^m}{R}\tau +
\frac{\omega^m R}{\alpha^\prime}\sigma + i\sqrt{\frac{\alpha^\prime}{2}} \sum\limits_{n\neq 0} \frac{1}{n} [ \alpha^m_n e^{-2in(\tau -\sigma)} + {\tilde\alpha}^m_n e^{-2in(\tau +\sigma)} ]. \label{modescompact}
\end{gather} 
The sixteen compact bosons $H^I, I=1,..,16$ arising on the left-moving sector of heterotic string theory have similar mode expansion as of $X^m$.
The mode expansion of the fermions are given by
\begin{gather}
S_+^a = \sum\limits_{n=-\infty}^\infty ~ 
S_n^a e^{-2in(\tau + \sigma)},
\quad\quad
S_-^a = \sum\limits_{n=-\infty}^\infty ~ 
{\tilde S}_n^a e^{-2in(\tau - \sigma)}.
\end{gather}
The oscillators satisfy the following commutation relations
\begin{gather}
[\alpha_m^i , \alpha_n^j ] = m\delta_{m+n}\delta^{ij},
\quad\quad
\{ S_m^a , S_n^b \} = \delta_{m+n} \delta^{ab},
\end{gather}
and a similar set of equations for the left movers. The bosonic zero modes $x$ and $p$ satisfy
\begin{gather}
[x^i , p^j] = i\delta_{ij}.
\end{gather}

The light-cone gauge is given in terms of the light-cone coordinate,
\begin{gather}
X^\pm = \frac{1}{\sqrt{2}} ( X^0 \pm X^9 ),
\end{gather}
by setting the oscillators along one of these directions to zero,
\begin{gather}
X^+ = x^+ + \frac{1}{\sqrt{2}}
\left[ \alpha^\prime
(p^0 + \frac{q^9}{R})\tau + 
\frac{\omega^9 R}{\alpha^\prime}\sigma 
      \right].
\end{gather}
Using the vanishing of components of stress 
tensor $T_{++}=T_{--}=0$
one can solve the oscillators in the mode expansion of $X^-$ in terms 
of the transverse bosonic oscillators and the oscillators 
coming from the fermions.
For heterotic string theory in left-moving sector, the contributions in addition to those from transverse bosonic oscillators, will come from internal coordinates.
For convenience, we introduce
\begin{gather}
q^+_{R,L} = \frac{1}{\sqrt{2}} \left[
\alpha^\prime ( p^0 + \frac{q^9}{R}) \pm \frac{\omega^9 R}{\alpha^\prime}  \right].
\end{gather}
In this notation, the right-moving and left-moving 
components of $X^+$ can be written as
\begin{gather}
\partial_+ X^+ = \frac{1}{2} q_R^+,
\quad
\partial_- X^+ = \frac{1}{2} q_L^+.
\end{gather}
Similarly, the right-moving and left-moving components
of the transverse bosonic components are
\begin{gather}
\partial_+ X^i = \sqrt{2\alpha^\prime} 
\sum\limits_{n\neq 0} \alpha_n^i  e^{-2in(\tau + \sigma)} ,
\quad
\partial_- X^i = \sqrt{2\alpha^\prime} 
\sum\limits_{n\neq 0} {\tilde\alpha}_n^i  e^{-2in(\tau - \sigma)} .
\end{gather}
The two dimensional action can be written as
\begin{gather}
S = \frac{1}{2\pi\alpha^\prime} \int ~ d\sigma d\tau ~ ( 2 \partial_+X^i\partial_-X^i +  2\partial_+X^m\partial_-X^m + i\alpha^\prime S_+^\alpha
\partial_-S_+^\alpha + i\alpha^\prime {\tilde S}_-^\alpha
\partial_-{\tilde S}_-^\alpha ),
\end{gather}
where we write down the action for type II. For the heterotic string theory the left-moving fermionic piece will be replaced by that of sixteen internal bosonic coordinates.

\subsection{Symmetries of the hologram}

Now we discuss the symmetries of the conformal field theory in this light-cone gauge. There is a global spin(4)-symmetry, which is the relic of spin(8)-symmetry after four-torus compactification and eight global supersymmetries generated by $(Q^a_0, Q^{\dot a}_0)$ in each sectors (eight global supersymmetries in the right-moving sector for heterotic theory). In addition, there is conformal symmetry in both the sectors and a chiral affine spin(8)-symmetry in both the sectors of type II strings (in the right-moving sector of heterotic string). In the following we write down the generators and the associated algebra.  

\subsubsection*{spin(4)-symmetry}
The generators of global non-chiral spin(4)-symmetry are similar to the standard Lorentz generators of spin(8)-symmetry and are given by,
\begin{gather}
J^{ij} = L^{ij} + E^{ij} + K_0^{ij} + {\tilde E}^{ij} + 
{\tilde K}_0^{ij}
\label{globalspin4}\end{gather}
where $i,j=1,..,4$.
The various components of this expression are given by
\begin{gather}
L^{ij} = (x^ip^j-x^jp^i),\\
E^{ij} = -i\sum\limits_{n>0} \frac{1}{n} (\alpha_{-n}^i\alpha_n^j - \alpha_{-n}^j\alpha_n^i ),\\
K_0^{ij} = -\frac{i}{4}\sum\limits_{n>0} ~ S^a_{-n}\gamma^{ij}_{ab} S_n^b,
\quad
\gamma^{ij}_{ab} = (1/2)[\gamma^i, (\gamma^j)^T]_{ab},
\label{affinecurrent}
\end{gather}
where $(\gamma^i)^T$ is the transpose of $\gamma^i_{a{\dot a}}$, 
$a$ and ${\dot a}$ denote $8_s$ and $8_c$ Majorana-Weyl representation of spin(8); $i$ denotes $8_v$ of $SO(8)$ and 
$\gamma^i_{a{\dot a}}$ are Clebsch-Gordon coefficient among $8_v$, 
$8_s$ and $8_c$. We use
\begin{equation}
\begin{split}
\gamma^1 &= \epsilon\times\epsilon\times\epsilon,
\quad
\gamma^5 = \sigma_3\times\epsilon\times 1_2,
\quad
\gamma^2 = 1_2\times\sigma_1\times\epsilon,
\quad
\gamma^6 = \sigma_1\times\epsilon\times 1_2,\\
\gamma^3 &= 1_2\times\sigma_3\times\epsilon,
\quad
\gamma^7 = \epsilon\times 1_2 \times \sigma_3,
\quad
\gamma^4 = \epsilon\times 1_2 \times\sigma_1,
\quad
\gamma^8 = 1_2\times 1_2\times 1_2,\label{gamma}
\end{split}
\end{equation}
where $\epsilon=i\sigma_2$.
Similarly expressions of ${\tilde E}^{ij}$ and ${\tilde K}_0^{ij}$ can be obtained by replacing the oscillators in the above expressions by their left-moving counterpart. These generators can be thought of as associated with the global isometry of three-sphere horizon. In addition to this global non-chiral symmetry we have a number of chiral affine symmetries.

\subsubsection*{Conformal symmetry}
For the right-moving sector, we have a conformal symmetry generated by ${\cal L}_n$, the modes of the following operator
\begin{gather}
q_R^+ (\partial_+ X^-) = \frac{1}{2\alpha^\prime}(\partial_+X^i\partial_+X^i) + \frac{i}{2}S^a_+\partial_+S^a_+,
\end{gather}
and a similar set of generators of conformal symmetry in the left-moving sector of type II theory ( for heterotic theory
the fermions need to be replaced by internal bosons). 
The modes of energy-momentum tensor satisfy Virasoro algebra
given by
\begin{gather}
[{\cal L}_m,{\cal L}_n] = (m-n){\cal L}_{m+n} + \frac{c}{12}(m^3-m)\delta_{m+n},
\end{gather}
where the central charge $c=12p$.
There is a global $SL(2,R)\times SL(2,R)$ symmetry, generated by 
$({\cal L}_0,{\cal L}_1, {\cal L}_{-1})$ and 
$({\tilde{\cal L}}_0,{\tilde{\cal L}}_1, {\tilde{\cal L}}_{-1})$
which can be identified with isometry group of AdS$_3$.

\subsubsection*{Supersymmetry}
Both the sectors of type II theories (only right-moving sector of heterotic theory) have global supersymmetry generated by supercharges
which are zero modes of the following spin-3/2 supercurrent.
\begin{gather}
Q^{\dot a}(\sigma^+) = \frac{1}{\sqrt{q^+_R}}(\gamma^i)^{\dot a}_{~a} S^a_+\partial_+X^i, \label{supercharge}
\end{gather}
where $\gamma$-matrices are given in (\ref{gamma}).
In addition, the modes of supercurrent satisfy the following affine algebra\cite{Green:1980zg},
\begin{gather}
\{Q_m^{\dot a} , Q_n^{\dot b}\} = 
2\delta^{{\dot a}{\dot b}}{\cal L}_{m+n} +
(m-n)(\gamma^{ij})^{{\dot a}{\dot b}} K^{ij}_{m+n} +{\tilde c}(m^2 - \frac{1}{4})\delta^{{\dot a}{\dot b}}\delta_{m+n},
\end{gather}
where ${\tilde c}$ is a constant.
The commutation relations between these modes of supercurrents and Virasoro generators are given by
\begin{gather}
[{\cal L}_m , Q^{\dot a}_n ] = (\frac{1}{2}m - n) Q^{\dot a}_{m+n}.
\end{gather}

\subsubsection*{spin(8)-symmetry}
In addition to the global spin(4)-symmetry, both the sectors of type II theories (only right-moving sector of heterotic theory) have an affine spin(8)-symmetry generated by the 
spin one currents
\begin{gather}
K^{ij}(\sigma^+) = -\frac{i}{4} S_+^a \gamma^{ij}_{ab}S^b_+.
\end{gather}
For heterotic theory one has a similar affine current algebra associated with $E_8\times E_8$.
The modes of affine spin(8) current satisfy the following Kac-Moody algebra
\begin{gather}
[ K_m^{ij}, K_n^{kl} ] = i(\delta^{ik} K^{ji}_{m+n} - \delta^{il} K^{jk}_{m+n}) + \frac{\tilde k}{2}(m-n)\delta_{m+n}(\delta^{ik}\delta^{jl}
-\delta^{jk}\delta^{il}).
\end{gather}
The modes of affine currents have the following commutator with Virasoro generators,
\begin{gather}
[{\cal L}_m , K^{ij}_n ] = -n K^{ij}_{m+n}.
\end{gather}
As discussed in \cite{Dabh}
the commutators of modes of affine current and supercharges do not close
\begin{gather}
[K^{ij}_m , Q^{\dot a}_n] = - \frac{i}{4}(\gamma^{ij}\gamma^k)_{a{\dot a}} \sum\limits_r ~ S^a_{-r}\alpha^k_{m+n+r},
\end{gather}
which is consistent with the fact that 
it is not possible to have a $(8,0)$ superconformal 
algebra with a spin(8) R-current.

Though we cannot write an ${\cal N}=8$ superconformal algebra, as described in \cite{Dabh} one can write an ${\cal N}=2$ superconformal algebra by choosing a $U(1)$-current $J= K^{12} + K^{34} + K^{56} + K^{78}$ and forming two linear combinations $G^\pm$ of eight supercharges. Since the construction is identical as in \cite{Dabh}, we are not repeating that. Similarly, it is possible to construct an ${\cal N}=4$ superconformal algebra along the same line.

\subsection{Internal light-cone gauge:}
As we will see in the next section the symmetry algebra in the light-cone gauge discussed above, does not match with that on the bulk side.
Therefore, following \cite{Dabh}, we modify the light-cone gauge so that the algebra matches with that in its holographic dual.

\subsubsection*{Left-moving sector (heterotic string)}
We begin with the left-moving sector of heterotic theory. 
We pair up the boson along compact direction $X^9$ with one of the internal boson 
$H^1$ (that generates an $SU(2) \subset E_8\times E_8$) through the following coordinates
\begin{gather}
Y=\frac{1}{\sqrt 2} (X_L^9 + H^1),
\quad\quad
{\bar Y}=\frac{1}{\sqrt 2} (X_L^9 - H^1).
\end{gather}
The light-cone gauge is given in terms of the following light-cone coordinates
\begin{gather}
X^{\pm} = \frac{1}{\sqrt{2}}(X^0 \pm Y),
\end{gather}
as
\begin{gather}
\partial_-X^+ = \frac{1}{2}q_L^+ \\
q_L^+ = \frac{1}{\sqrt{2}} [\alpha^\prime ( p^0  + \frac{1}{\sqrt{2}}(p^9.R^9 + p'.R')) - \frac{1}{\sqrt{2}\alpha^\prime}(\frac{q^9}{R^9} + \frac{q'}{R'})],
\end{gather}
where we use $(p^9, q^9; R^9)$ and $(p', q'; R')$ to be momentum and winding and radius associated with $X^9$ and $H^1$ respectively. Apart from light-cone coordinates we are left with 24 bosons given by
\begin{equation}
\{ \phi^i \} = \{ X^1, \dots,X^8, {\bar Y}, H^2, \dots, H^{16} \},
\end{equation}
which we collectively call $\phi^i$.
Using the Virasoro constraint one can solve for other light-cone coordinate $X^-$ and given by 
\begin{gather}
q^+\alpha_n^- = \frac{1}{2} \sum\limits_{i=1}^{24}\sum\limits_n \alpha^i_{n-m}\alpha^i_m = {\cal L}_n. \label{virasoroL}
\end{gather}

Now we consider how this gauge modifies the symmetry algebra arising in the left-moving sector of heterotic string. 
Other symmetries will remain the same, except, this gauge spoils the 
$SU(2)\subset E_8\times E_8$ algebra associated with the boson $H^1$. The currents are given by 
$\partial_-H^1$, $e^{\pm i\sqrt{2} H^1}$. The mode expansion of the first one becomes
\begin{gather} 
{\tilde j}_n = \frac{1}{4} q^+ \delta_{n,0} - \frac{1}{2}{\tilde\alpha}_n^- - \frac{1}{\sqrt{2}}{\tilde\alpha}_n^{\bar Y}.
\label{jn}\end{gather}
In this gauge we have the commutator
\begin{gather}
[{\tilde{\cal L}}_m , {\tilde j}_n ] 
= - \frac{1}{2 q_L^+} [{\tilde{\cal L}}_m, {\tilde{\cal L}}_n ] 
- \frac{1}{2\sqrt{2}} \sum\limits_k 
[ {\tilde{\cal L}}_m , {\tilde\alpha}^{\bar Y}_n ].
\end{gather}
Using the expression given in (\ref{virasoroL}) we can see
that it is different from what is expected for modes of a spin one current, {\it i.e.} $[{\tilde{\cal L}}_m, {\tilde j}_n ]= -n {\tilde j}_{m+n}$. In particular 
${\tilde j}_0$ has the commutation same as that obtained in the sigma model picture.

\subsubsection*{Right-moving sector}
In order to describe the special gauge in the right-moving sector we will bosonize the fermions in the Green-Schwarz formulation and use a linear combination of them. This can be thought of as related to eight world-sheet fermions $\psi^1, ...,\psi^8$ in the R-NS formulation in the following way. Bosonizing the latter into four bosons,
\begin{gather}
-\partial\theta_i = \psi^{2i-1}\psi^{2i},
\quad\quad i =1,2,3,4.
\end{gather}
one connect them to the fermions in Green-Schwarz formulation 
\begin{gather}
S^{(2i-1)} S^{(2i)}= -\partial \sigma_i,
\quad\quad
i=1,2,3,4, \label{gsfermion0}
\end{gather}
through the relation
\begin{gather}
\sigma_1 = (1/2)[(\theta_1 + \theta_2) + (\theta_3 + \theta_4)],
\quad\quad
\sigma_2 = (1/2)[(\theta_1 + \theta_2) - (\theta_3 + \theta_4)],\\
\sigma_3 = (1/2)[(\theta_1 - \theta_2) + (\theta_3 - \theta_4)],
\quad\quad
\sigma_2 = (1/2)[(\theta_1 - \theta_2) - (\theta_3 - \theta_4)].
\end{gather}

The $SO(8)$ currents in terms of GS fermions are given
in (\ref{affinecurrent}). Substituting the expressions of GS fermions in terms of $\theta_i$'s and using
$(\gamma^i)_{a{\dot a}}$ matrices given in (\ref{gamma}),
we obtain $SO(4)$ currents splitted into two commuting sets of $SU(2)$ currents,
\begin{gather}
K^{12} - K^{34} = i\partial(\theta_3 + \theta_4 ),
\quad\quad
( K^{31} - K^{24} ) \mp i (K^{23} - K^{14}) =  
\sqrt{2} e^{\pm i(\theta_3 + \theta_4 )} ,\label{so4current1}\\
K^{12} + K^{34} = i\partial(\theta_3 - \theta_4 ),
\quad\quad
(K^{31} + K^{24}) \mp i (K^{23} + K^{14})  = \sqrt{2} 
e^{\pm i(\theta_3 - \theta_4 )} \label{so4current2}.
\end{gather}

As in the left-moving sector we combine $X^9$ and $X^\theta = \frac{1}{\sqrt{2}} (\theta_3 + \theta_4)$ to define new coordinates,
\begin{gather}
Y = \frac{1}{\sqrt{2}} ( X^9 + X^\theta ),
\quad\quad
{\bar Y} = \frac{1}{\sqrt{2}} ( X^9 - X^\theta ),
\end{gather}
and state the light-cone gauge in terms of them as
\begin{gather}
\partial_+ X_R^+ = \frac{1}{2} q_R^+,\quad
{\text where}, \quad
X_R^\pm = \frac{1}{\sqrt{2}} ( X^0 \pm Y ),
\end{gather}
and
\begin{gather}
q_R^+ = \frac{1}{\sqrt{2}} \left[ \alpha^\prime (p^0 +  \frac{1}{\sqrt{2}}(p^9.R^9 + p^\theta.R^\theta) + \frac{1}{\sqrt{2}\alpha^\prime} (\frac{q^9}{R^9} + \frac{q^\theta}{R^\theta})\right].
\end{gather}

In addition, we fermionize the other boson $\frac{1}{\sqrt{2}}(\theta_3 - \theta_4)$ into a pair of world-sheet fermions through
\begin{equation}
(K^{12} + K^{34}) = i \partial(\theta_3 - \theta_4) =  \sqrt{2} \chi^3 \chi^4,
\end{equation}
and combine one of them, say, $\chi^3$ with $\psi^9$
as
\begin{gather}
\psi^Y = \frac{1}{\sqrt{2}}( \psi^9 + \chi^3),
\quad\quad
\psi^{\bar Y} = \frac{1}{\sqrt{2}}( \psi^9 - \chi^3),
\end{gather}
and impose light-cone gauge as
\begin{gather}
\psi^+ = 0, 
\quad\quad 
\text{where},
\quad\quad \psi^\pm = \frac{1}{\sqrt{2}} ( \psi^0 \pm \psi^Y ).
\end{gather}
Apart from the light-cone coordinates we have the following transverse fields:
\begin{equation}
\{ X^1, \dots X^8 , {\bar Y} , \chi^4 , \psi^1, \dots, \psi^4, \psi^{\bar Y} \}.
\end{equation}
In this gauge solving Virasoro and supersymmetry constraints we obtain 
\begin{gather}
{\cal L}_n = q_R^+\alpha_n^- , 
\quad\quad
G_n = q_R^+ \psi_n^- , 
\end{gather}
which are the expressions for modes of Virasoro and supersymmetry generators.

This gauge modifies the properties of the $SO(4)$ currents as follows.
The modes of
$N^3 = (K^{12} - K^{34})$ in (\ref{so4current1}) are given by
\begin{gather}
N^3_n = \frac{1}{4} q_R^+ \delta_{n0} - \frac{1}{2}\alpha_n^- - \frac{1}{\sqrt{2}}\alpha_n^{\bar Y}.
\label{kn}\end{gather}
Comparing this with the expression of ${\tilde j}_n$, one of the generator of $SU(2)$ in the left-moving sector, given in (\ref{jn}) we see that $N^3$ does not have the correct conformal dimension and therefore the associated affine current is absent. However, we can consider the corresponding global symmetry generated by zero mode
of the current, which satisfies
\begin{gather}
[{\cal L}_m , N^3_0 ] = m N^3_0. 
\label{LN3commutatorhologram}
\end{gather}
Since it is $N_3^0$ that enters the generator of global $SU(2)\subset spin(4)$ associated with (\ref{so4current1}) given in
(\ref{globalspin4}) we can say this gauge modifies the commutator of generators of
$SU(2)\subset spin(4)$ with $SL(2, R)$ generators in the above manner. 

In order to match with the algebra obtained on the bulk side, which we will describe latter, for the left moving sector of type II theory we choose $X^\theta = \frac{1}{\sqrt{2}} (\theta_3 - \theta_4)$ to define new coordinate $Y$ and ${\bar Y}$ and carry out the gauge fixing. As a result the generators in (\ref{so4current2}) of the other $SU(2)$ will have unusual commutator with Virasoro generators coming from the left moving sector. 


In this gauge, computation of the other components of currents is difficult and construction of the supercharges needs further work. Nevertheless, since the original world-sheet theory has eight supercharges in each sector ( only in right-moving sector for heterotic theory) and we made a special gauge choice it is possible to write down the supersymmetry generators which will be modified from their form in (\ref{supercharge}) by some compensating gauge transformation.

\section{String theory dual}

In this section we discuss the sigma model that describes
string theory on the near horizon geometry of macroscopic
string which is $AdS_3\times S^3\times T^4$. The geometry
has an AdS$_3$ piece and therefore the sigma model should
reflect the associated isometry $SL(2,{\bf R})\times SL(2,{\bf R})$. 
There are extensive studies about superstring theory on 
AdS$_3$ in literature \cite{seiberg1,seiberg2, beren,hikida}
and it turns out that the sigma model is described
by an $SL(2,{\bf R})_k$ WZW model where $k$ is the level of
the algebra and is related to the size of AdS$_3$ space.
Our sigma model, then, contains a supersymmetric $SL(2,{\bf R})_k$
WZW model for each sector and a bosonic $SL(2,{\bf R})_k$ WZW
model for the left-moving sector of Heterotic string.

The second piece comes from considering the isometry 
of $T^4$ and is given by supersymmetric {\cal N}=1 $U(1)^4$
SCFT. That has a free field theoretic realization in terms of four compact bosons and their fermionic partners. For the left-moving sector of the heterotic string we
need a bosonic $U(1)^4$ CFT and a bosonic $U(1)^{16}$ WZW
model to generate $E_8\times E_8$ symmetry.

Now let us count the central charges of the supersymmetric sector. A supersymmetric $SL(2,{\bf R})_k$ 
WZW model contributes $\frac{9}{2} + \frac{6}{k}$ to the sigma model world-sheet central charge. As shown in \cite{seiberg1} this gives 
rise to a super-affine
$SL(2,{\bf R})$ algebra in the boundary theory with a central charge $c=6kp$,
where $p$ is the winding. Identifying this with the right-moving
transverse superstring which has a space-time central charge $12p$ we get
$k=2$. Therefore the contribution of $SL(2,{\bf R})_2$ WZW model to the central charge is $c= 15/2$. The {\cal N}=1 $U(1)^4$ SCFT contributes $c=6$ to the central charge. To get 15 we need an extra 3/2 contribution to the central
charge which can naturally be obtained by utilizing three fermions 
$\xi_a,\;a=1,2,3$\footnote{The central 3/2 can also be obtained by a supersymmetric
WZW $SU(2)$ model in level of 2. We note, however, that with this choice the consistent theory 
can only support four supercharges. In fact this model has been studied 
in \cite{seiberg1} which describes $AdS_3\times S^3$ coming from D1/D5 system.}. 
So the sigma model is $SL(2,R)_2 \times\{\xi_1,\xi_2,\xi_3\} \times U(1)^4$ WZW model one for each sector of type II theory. This sigma model has also been proposed in the context of small black holes in \cite{giveon} where the author 
has found this particular model from different considerations in the context of five dimensional Heterotic 
small black hole. We note, however, that from the way we found this background it is evident that
it can describe both type II and heterotic small black string in six dimensions.

For left-moving sector of heterotic theory the counting goes as follows. The level of bosonic $SL(2,{\bf R})$ $k_b$ should be related to $k$ on the right-moving sector through $\sqrt{k_b/k} = \sqrt{2}$ \cite{giveon}. So
$k_b=4$ and contributes $c=3\frac{k_b}{k_b-2}=6$ to the central charge.
The $U(1)^4$ and the $U(1)^{16}$ contributes $c=20$. Since total central charge is already $c=26$ we cannot add any more degrees of freedom. The resulting sigma model is then bosonic $SL(2,{\bf R})_4\times\times U(1)^4\times U(1)^{16}$ CFT.

Now we consider the field content of the SCFT. 
For reasons explained in the sequel 
we follow \cite{Dabh,hikida} and split
the $SL(2,{\bf R})_2$ WZW model in two parts: an $({SL(2,{\bf R})}/{U(1)})$ and an
$U(1)$ which we parametrize by the bosons $(\tau,\rho)$ and
$(\theta)$ respectively, where the compact boson $\theta$ 
is at the free fermion radius. The corresponding fermions are given as
$(\psi_\tau, \psi_\rho)$ and $(\psi_\theta)$.
The $U(1)^4$ WZW is realized by
four compact bosons $Y^i,i=1,2,3,4$ and four fermions $\lambda_i$. On top of them we have three fermions
$\xi_a$'s.

This SCFT furnishes representation of ${\cal N}=1$
world-sheet superconformal algebra whose elements are given as
follows\cite{Dabh}. The world-sheet energy momentum tensor is given
by
\begin{equation}
\begin{split}
T = ~ - &\frac{1}{2} [ ~ (\partial\rho)^2 + \partial^2\rho +
(\partial\tau)^2 + (\partial\theta)^2 + (\partial
Y^i)^2 ~ ] \\
 - &\frac{1}{2} [ ~ \psi_\rho\partial \psi_\rho + \psi_\tau\partial
\psi_\tau + \psi_\theta\partial \psi_\theta + \xi^a \partial \xi_a +
\lambda_i\partial \lambda_i ~ ]. \label{emtensor}
\end{split}
\end{equation}
From (\ref{emtensor}) one can see $\rho$ is a Liouville-like field
while other fields are free. Similarly the supercurrent is given by
\begin{equation}
G  = i \left( ~ \psi_\tau\partial\tau +
\psi_\rho\partial\rho +
\partial\psi_\rho + \psi_\theta\partial\theta  +
\lambda_i\partial Y^i-\xi^1\xi^2\xi^3~ \right).
\label{supercurrent}
\end{equation}
This type of sigma model has been discussed in various
contexts \cite{hikida,beren,giveon}.

Similarly we realize the bosonic CFT describing the world-sheet dynamics of sigma model for left-moving sector of heterotic string by splitting the bosonic $SL(2,{\bf R})$ into two pieces: an $({SL(2,{\bf R})}/{U(1)})$ and an
$U(1)$ which we parametrize by the bosons $(\tilde\tau,\tilde\rho)$ and
$(\tilde X)$ respectively, where the compact boson $\tilde X$ 
is at the self-dual radius and we put tilde to indicate that they belong to left-moving sector. In addition we have twenty compact bosons given by
$\tilde Y^m; m=1,2,3,4$ and $\phi^I; I=1,..,16$.

\subsection{Symmetries of the sigma model}

In this subsection we discuss symmetries of the SCFT described above. For type II strings we get two SCFTs one for each sector. For heterotic string this describes the right-moving sector only and the left-moving sector will be discussed in the next subsection.

\subsubsection*{$Sl(2,{\bf R})$ symmetry}

We begin with the affine $SL(2, {\bf R})$ algebra that arises
as symmetry of AdS$_3$ geometry in the near horizon and reflected 
in the hologram through the Virasoro algebra discussed in the last section. The world-sheet operator associated with the generators
can be constructed as follows. Following \cite{Dabh,hikida} we write
down the operator in (-1)-picture as,
\begin{equation}
J_n^{(-1)}(z) = e^{2n(\tau + i\theta)} (\frac{1}{2}\psi_\tau -
n \psi_\rho).
\end{equation}
Then we use the relation
\begin{equation}
G(z)J^{(-1)}(w) = \frac{1}{z-w} J^{(0)}(w),
\label{pixchange}\end{equation}
to obtain the operator in the (0)-picture which has the following expression,
\begin{equation}
J_n^{(0)}(z) = e^{2n(\tau + i\theta)}(\frac{1}{2}\partial\tau - n
\partial\rho + 2 n^2 (\psi_\tau + i\psi_\theta)\psi_\rho - i n
\psi_\tau\psi_\theta ).
\end{equation}
Using these operators one can define the generators of
space time $SL(2, {\bf R})$ algebra
\begin{equation}
 {\mathcal L}_m = \oint dz J_m^{(0)}(z) , \label{sl2current}
\end{equation}
which satisfies the Virasoro algebra
\begin{equation}
[ ~ {\mathcal L}_m , ~ {\mathcal L}_n ] ~ = ~ (m-n) ~{\mathcal
L}_{m+n} + \frac{c}{12} m (m^2-1) \delta_{m+n,0}.
\end{equation}
As we see these operators generate an affine $SL(2, {\bf R})$ algebra which matches nicely with Virasoro algebra obtained on the hologram.
For
type II strings we get two copies of such $SL(2,{\bf R})$ one
from each chirality sector while for heterotic we get one from right-moving sector
of these generators $({\mathcal L}_0 , {\mathcal L}_1 , {\mathcal
L}_{-1} )$ generate the global $SL(2, {\bf R})$ symmetry group.

\subsubsection*{$SU(2)$ symmetry}

Comparing with \cite{Dabh} one notices that there 
is actually an $SU(2)$ current, which
in their case, corresponds to $SU(2)$ symmetry ensuing
from the isometry of $S^2$. The construction goes as follows. 
One fermionizes the compact boson $\theta$ at free fermion 
radius as $e^{\pm\ i \theta} = \frac{1}{\sqrt{2}}(
\psi_\theta^1 \pm \psi_\theta^2 )$. Then identifying 
$\psi_\theta^3\equiv \psi_\theta$ one constructs the
set of currents 
$(\psi_\theta^1 , \psi_\theta^2 ,\psi_\theta^3)$.
We identify them with the generator of $SU(2)$ algebra
in (-1)-picture, 
\begin{equation}
{N}^{I(-1)}_0 (z) = \psi_\theta^I , 
\quad\quad 
I=1,2,3,
\end{equation} 
Using (\ref{pixchange}) we get operators in the (0)-picture,
which gives rise to the following space-time generators 
for global 
$SU(2)$
\begin{equation}
{\mathcal N}^3 = \oint dz ~ (i\partial\theta)  ,
\quad\quad 
{\mathcal N}^\pm = \oint dz  ~ \mp
i ( \psi^3\psi^\pm ).
\end{equation}
As shown in \cite{Dabh} these generators do not have appropriate commutation relation with Virasoro generators. 
In particular, $[{\mathcal L}_m , {\mathcal N}^3 ] = m{\mathcal L}_m$ 
while the commutator should vanish. Nevertheless, if we compare 
with the hologram we can see that this is precisely same as
the commutator obtained there at the special
light-cone gauge.  In view of that, the $SU(2)$ generated by
${\mathcal N}^I$'s is the one that is reflected in the hologram.

Similarly on the left moving sector of type II theories, the potential generators of ${\widetilde SU(2)}$ will have this type of commutator with ${\widetilde SL(2,{\bf R})}$. On the other hand, $SU(2)$ generators commute with ${\widetilde SL(2,{\bf R})}$ which matches with what we got on the hologram side.

At this point we note that, as explained in \cite{Dabh}
since $\theta$ itself is engaged in the construction
of ${\mathcal N}^I$ it is not clear how to obtain its affine
extension. The standard procedure of multiplying by the 
factor $e^{2n(\tau + i\theta)}$  does not work. It has been proposed in \cite{Dabh} to construct the
affine current operators along the line of \cite{seiberg2}.
\subsubsection*{$U(1)^4$ symmetry}
One realizes the space-time affine $U(1)^4$ symmetry associated with the $T^4$ piece of the near horizon geometry as follows. The generators in (-1)-picture, can be obtained as \cite{Dabh,hikida}
\begin{equation}
P^{i(-1)}_n (z) = e^{2n(\tau + i\theta)}\lambda^i , i=1,..,4.
\end{equation} 
Then using the relation (\ref{pixchange})
and computing the operator product with $G(z)$ we obtain
the generators in the (0)-picture,
\begin{equation}
P^{i(0)}_n (z) = i e^{2n(\tau + i\theta)} [
\partial Y^i - 2n(\psi_\tau + i\psi_\theta)\lambda^i ]
\end{equation}
From these the generators of space-time $U(1)^4$ can
be constructed as
\begin{equation}
{\mathcal P}^i_n ~ = ~ \oint ~ dz ~ P^{i(0)}(z).
\end{equation}

\subsubsection*{Supersymmetry}

In order to construct the space-time supersymmetry generator we
use the standard method of the enhancement of 
${\cal N}=1$ superconformal symmetry of the world-sheet theory to 
${\cal N}=2$ superconformal. This requires construction of an 
$U(1)_R$ current in the world-sheet theory that splits the 
supercurrent $G$ in (\ref{supercurrent}) into two parts, with positive and negative charges \cite{fms}. 
We consider the ${\cal N}=1$ superconformal 
algebra generated by the world-sheet energy-momentum tensor (\ref{emtensor})
and supercurrent (\ref{supercurrent}) and define the $U(1)_R$ current in the following way. 
We can introduce the bosons $(H, H_1)$ given by
\begin{gather}
\partial H = - \psi_\rho ~ \psi_\theta + \partial\theta,
\label{R1} \\
\partial H_1 =\frac{1}{3} \psi_\tau(\xi^1+ \xi^2+\xi^3). \label{R2}
\end{gather}

These specific forms of the boson $H$ in (\ref{R1},\ref{R2}) are
required to avoid the possible double-pole arising in the operator
product of $U(1)_R$ current and the world-sheet supercurrent $G(z)$
(\ref{supercurrent}).
We introduce two more bosons by bosonizing the fermions 
corresponding to $T^4$ direction and write
\begin{equation}
 \partial H_2 = \lambda^1\lambda^2, \quad\quad
\partial H_3 = \lambda^3\lambda^4 \label{u4boson}.
\end{equation}
Then we choose the $U(1)$ R-symmetry current to be
\begin{equation}
J=J^1+J^2 , \quad\quad J^1=i\partial H, \quad\quad J^2=i(\partial
H_1+\partial H_2 +\partial H_3). \label{Rcurrent}
\end{equation}
This {\cal N}=2 superconformal algebra is very similar to the
one discussed in appendix of \cite{seiberg1} where this theory has a different interpretation.

Using the set of bosons given in (\ref{R1},\ref{R2},\ref{u4boson})
we construct the supercharges as follows
\cite{seiberg1}. We define $SO(4)$ spinors (associated with rotation
among $Y^i$'s ) $S_\alpha$ and $S_{\bar{\alpha}}$ which are in $(2,
0)$ and $(0, \bar 2)$ respectively of $SO(4)\approx SU(2)\times
SU(2)$ as
\begin{gather}
S_\alpha = e^{\pm\frac{i}{2}(H_2+H_3)}, \quad\quad S_{\bar{\alpha}}
 = e^{\mp\frac{i}{2}(H_2-H_3)}.
\end{gather}
Then we consider the $SU(2)$ spinors
\begin{gather}
S^{(\pm)}_\alpha = e^{-\frac{\phi}{2}} ~ e^{\pm\frac{i}{2}H} ~
e^{\frac{i}{2} H_1} ~ S_\alpha, \quad\quad {\bar
S}^{(\pm)}_{\bar\alpha} = e^{-\frac{\phi}{2}} ~ e^{\pm\frac{i}{2}H}
~ e^{-\frac{i}{2}H_1} ~ S_{\bar\alpha},
\end{gather}
where $\phi$ is the bosonized ghost. The global supercharges can be
given in terms of the above operators as
\begin{gather}
Q^{(a)}_\alpha = \oint dz S^{(a)}_\alpha (z), \quad\quad {\bar
Q}^{(a)}_{\bar\alpha} = \oint dz {\bar S}^{(a)}_{\bar\alpha} (z) ,
\quad a = \pm .
\end{gather} 
Counting the different components we get altogether 8
supercharges. Let us now consider the anti-commutators of the following
supercharges
\bea
\{ ~Q^{(+)}_\alpha , ~Q^{(-)}_\beta \} &=&\frac{1}{3} \sigma^1_{\alpha\beta} ~(3
{\mathcal L}_0 + {\tilde{\mathcal N}}^3_0+{\tilde{\mathcal N}}^2_0+ {\tilde{\mathcal N}}^1_0), 
\cr 
\{ ~{\bar
Q}^{(+)}_{\bar\alpha}  ~{\bar Q}^{(-)}_{\bar\beta} \} &=&\frac{1}{3}
\sigma^1_{{\bar\alpha}{\bar\beta}} ~( 3{\mathcal L}_0 - {\tilde{\mathcal
N}}^3_0-{\tilde{\mathcal N}}^2_0- {\tilde{\mathcal N}}^1_0).
\label{yyy}
\eea
Here ${\cal \tilde{N}}^I_n$ are defined by
\begin{equation}
{\tilde{\mathcal N}}^I_n = \oint dz {\tilde N}^{I(0)}_n(z),
\end{equation}
where 
\begin{gather}
{\tilde N}^{I(0)}_n (z) = -  e^{2n(\tau + i\theta)} \left(\frac{i}{2}\epsilon^I_{\hphantom{I}JK}\xi^J(z)\xi^k(z)
+2n(\psi_\tau+i\psi_\theta)\xi^I\right). \label{M-affine}
\end{gather}
Note that this generators satisfy a $SU(2)$ algebra as follows
\begin{equation}
\begin{split}
[~{\tilde{\mathcal N}}^I_m , ~ {\tilde{\mathcal N}}^J_n ~] 
~&= ~ if^{IJ}_K ~
{\tilde{\mathcal N}}^K_{m+n} + ~ p ~ m ~ \delta_{m+n,0} , \\
[~{\mathcal L}_m , ~ {\tilde{\mathcal N}}^I_n ~] ~&= ~-n ~{\tilde{\mathcal
N}}^I_{m+n}, 
\end{split}
\end{equation} 
where $I=1,2,3$ and $m,n \in {\bf Z}$. One could also define ${\tilde{\mathcal N}}^I_n$ in (-1) picture by
utilizing 
${\tilde N}^{I(-1)}_n(z)=e^{2n(\tau + i\theta)}\xi^I , I=1,2,3$.

The algebra (\ref{yyy}) is same as that obtained in the appendix B of
\cite{seiberg1} as we have used the same
construction in terms of different variables\cite{Dabh, hikida}. In fact the resulting $SU(2)$ can be
interpreted as $SU(2)_R$ subgroup of the $SO(4)$ isometry of $S^3$.

As the commutator $[{\mathcal L}_0 , Q]$ vanishes we 
identify the eight supercharges as corresponding to the Ramond
sector of the space-time superconformal algebra. In that case, the
right hand side of the commutators are not quite correct because
of the presence of ${\tilde{\mathcal N}}^I$ on right hand side.
One resolution of this puzzle is to restrict to the representation 
of the superconformal algebra that belongs to the zero eigenvalue of ${\tilde{\mathcal N}}^I$. 
Then due to non-vanishing commutators with zero and non-zero modes of other components of the currents the subspace should belong to zero eigenvalue of  ${\tilde{\mathcal N}}^I_n$. This might be related to the
observation of \cite{giveon} where the authors have shown that approaching the hear horizon we 
will lose some part of the symmetries.

The other anti-commutators are similar as those obtained in
\cite{seiberg1,Dabh}
\begin{equation}
\{ ~Q^{(+)}_\alpha , ~{\bar Q}^{(-)}_{\bar\beta} \}  =~
\delta_{{\alpha}{\bar\beta}} ~{\mathcal P}_1 +
\sigma^3_{{\alpha}{\bar\beta}} ~{\mathcal P}_2 +
\sigma^3_{{\alpha}{\bar\beta}} ~{\mathcal P}_3, +
i\sigma^2_{{\alpha}{\bar\beta}} ~{\mathcal P}_4, ,
\end{equation}
where $\sigma^1$, $\sigma^2$  and $\sigma^3$ are Pauli spin
matrices. 

In order to compare this algebra with the one obtained
on the hologram side we need to set ${\mathcal P}_i$'s equal to
zero. Then using the symplectic-Majorana Weyl components one can rewrite the algebra in a closed form which resembles the one obtained on the hologram side restricted to zero modes.

So far we obtain the zero mode of the supercharges in Ramond sector. That we have a Virasoro algebra indicates the fact that these supercharges can be extended into a superconformal algebra. One may use the commutator
\begin{equation}
[{\cal L}_m , Q_n^a ] = (\frac{m}{2} - n ) Q_{m+n}^a,
\end{equation}
to construct the higher modes of the supercurrent. 

We have discussed the Ramond sector of the space-time theory whose global symmetries match with the hologram. If ${\cal N}=2$ affine superconformal algebra can be realized spectral flow can lead to NS sector of the space-time theory\cite{seiberg1}. In this context it may be interesting to get the explicit superconformal symmetry from the near horizon geometry along the line of \cite{strominger} and compare this with the symmetry obtained in this sigma model.

\subsection{Left-moving sector of heterotic string}
In this subsection we discuss the symmetries of the bosonic CFT which described the world-sheet dynamics of left-moving sector of the heterotic string.
\subsubsection*{$SL(2,{\bf R})$ symmetry}
The form of the currents generating the $SL(2,{\bf R})$ symmetry is very similar to their right-moving counterparts. In order to have an affine extension of $SL(2,{\bf R})$ we need to borrow one compact boson $\tilde\theta$ from the sixteen compact boson realizing the $E_8\times E_8$ part. The currents, then are given by
\begin{gather}
J_n({\bar z}) = e^{(n/\sqrt{2})(\tilde\tau + i\tilde\theta)} {\bar\partial} ( \tilde\tau - (n/\sqrt{2})\tilde\rho)({\bar z}).
\end{gather}
The corresponding generators satisfy the Virasoro algebra on the left-moving sector.

\subsubsection*{$SU(2)$ symmetry}
The world-sheet operators associated with the bosonic $SU(2)$ generators can be constructed from the 
compact boson $\tilde X$ at the self-dual radius as follows
\begin{gather}
M^3({\bar z}) = i\bar\partial {\tilde X}({\bar z}),
\quad\quad
M^{\pm}({\bar z}) = e^{\pm i{\sqrt{2}\tilde X}}.
\end{gather}
The generators of space-time affine $SU(2)$, then can be constructed as
\begin{gather}
{\mathcal M}^I_n = \oint d{\bar z} e^{(n/\sqrt{2})(\tilde\tau + i\tilde\theta)} T^I({\bar z}),
\quad\quad I =1,2,3.
\end{gather}
At this point it may be mentioned that unlike type II the SU(2) coming from the left moving sector do have appropriate commutation relation with Virasoro generators while an $SU(2)\subset E_8\times E_8$ is spoiled in this case.

\subsubsection*{$E_8\times E_8$ symmetry}
The currents associated with the generators of $E_8\times E_8$ symmetry are given by
\begin{gather}
J_n({\bar z}) = \oint d{\bar z} V({\bar z}) e^{(n/\sqrt{2})(\tilde\tau + i\tilde\theta)} (\bar z),
\end{gather}
where $V$ is the standard dimension one operator. $V({\bar z})= V_a({\bar z})= e^{2i\phi^a}$ when the generator belongs to Cartan and $V({\bar z})=V_K({\bar z}) = e^{2i(K^a \phi^a)}$ when $K^a$ is a root vector. Note that we already have borrowed one of these compact bosons and so we can realize only $E_7\times E_8$ subalgebra. Moreover, like one of the $SU(2)_R$ the affine extensions does not have appropriate conformal dimension. 

\section{Discussions}

To summarize, in this note we reviewed the two dimensional conformal field theory of macroscopic string compactified 
on a four-torus which may serve as the hologram of the string theory on the near horizon geometry. On the bulk side we consider a sigma model and analyze the
possibility that it describes the string theory on the near horizon geometry. We compare the global symmetries coming from the both sides and find an agreement once the two dimensional CFT is considered in a special light-cone gauge. At this special light-cone gauge the global $SO(4) \sim SU(2)\times SU(2)$ symmetry is not manifest for type II theory while in heterotic theory one of the global $SU(2)$ alongwith an $SU(2)\subset E_8\times E_8$ are not manifest. In addition, on the bulk side we need to restrict ourselves to a subspace of possible states. We could not find the appropriate affine extensions of the symmetries and there may exist a different construction of affine generators as suggested in \cite{seiberg2}. 

Due to special light-cone gauge it is not possible to realize the full affine superconformal algebra. It would be very useful to find a formulation with ordinary light-cone gauge where this algebra can be manifested. In principle one would expect to have an ${\cal N}=8$ algebra but as discussed in \cite{strominger,Dabh} this does not have a linear realization and so it is not clear how to get this. In addition to the comparison of symmetries, other tests of these dualities can be comparison of states on the string theory side and the operators on the hologram side following the analysis\cite{maldacena,gaberdiel} that has been done in the context of D1/D5 system. We hope to come back with some of these issues in future.

In this paper we have only considered small black string in six dimensions. Since one would expect to
have small black string in any dimensions bigger than four, one might wonder how the 
consideration of this paper and those in \cite{Dabh} can be generalized for $D\geq 6$.
As far as the near horizon supergroups are concerned there are proposal for corresponding 
supergroup in any dimensions \cite{Dabh,strominger,kraus}, though it is not obvious how to 
generalize the consideration of the present paper (as well as \cite{Dabh}) to higher 
dimensions. In fact as far as the $SL(2,{\bf R})_2$ factor is concerned, it seems that the procedure
goes the same in any dimensions leading to a 15/2 contribution to the central charge as well as
having a $SU(2)$ symmetry enhancement. But it is not clear to us how the other factors of space-time
can be described by a well defined WZW model. This issue might also be related to the discussions
of replacing highly excited strings with geometries mentioned in \cite{giveon} which could be
very complicated for higher dimensions.  

\renewcommand{\theequation}{A.\arabic{equation}}
\setcounter{equation}{0} 

\section*{Acknowledgments}
We would like to thank F. Ardalan and A. Kumar for useful discussions. S.~M. thanks IPM, Tehran for hospitality where part of this work was done. This work is supported in part by Iranian TWAS chapter at ISMO.

\end{document}